\documentclass{ifacconf}

\usepackage{amsmath,amsfonts,amssymb}
\usepackage{graphicx}
\usepackage{natbib}        
\usepackage{epstopdf}

\newcommand{\be}{\begin{equation}}
\newcommand{\ee}{\end{equation}}
\newcommand{\bd}{\begin{displaymath}}
\newcommand{\ed}{\end{displaymath}}
\newcommand{\bea}{\begin{eqnarray}}
\newcommand{\eea}{\end{eqnarray}}
\newcommand{\R}{\mathbb{R}}
\newcommand{\C}{\mathbb{C}}

\newcommand{\w}{\omega}

\newcommand{\kp}{{(k)}}
\newcommand{\kmp}{{(k-1)}}

\newcommand\blfootnote[1]{%
  \begingroup
  \renewcommand\thefootnote{}\footnote{#1}%
  \addtocounter{footnote}{-1}%
  \endgroup
}

\DeclareMathOperator*{\argmax}{arg\,max}

\begin{document}
\begin{frontmatter}

\title{Transfer Function Estimation in System Identification Toolbox via Vector Fitting}

\author[First]{Ahmet Arda Ozdemir},
\author[First]{Suat Gumussoy}

\address[First]{MathWorks, \\
        3 Apple Hill Drive, Natick, MA 01760 \\
        \mbox{(e-mails: \{arda.ozdemir, suat.gumussoy\}@mathworks.com)}.}

\begin{abstract}
  This paper considers black- and grey-box continuous-time transfer function
  estimation from frequency response measurements. The first contribution is a
  bilinear mapping of the original problem from the imaginary axis onto the unit
  disk. This improves the numerics of the underlying Sanathanan-Koerner
  iterations and the more recent instrumental-variable iterations. Orthonormal
  rational basis functions on the unit disk are utilized. Each iteration step
  necessitates a minimal state-space realization with these basis functions. One
  such derivation is the second contribution. System identification with these
  basis functions yield zero-pole-gain models. The third contribution is an
  efficient method to express transfer function coefficient constraints in terms
  of the orthonormal rational basis functions. This allows for estimating
  transfer function models with arbitrary relative degrees (including improper
  models), along with other fixed and bounded parameter values. The algorithm is
  implemented in the \verb"tfest" function in System Identification Toolbox
  (Release 2016b, for use with MATLAB) for frequency domain data. Two examples
  are presented to demonstrate the algorithm performance.
\end{abstract}


\end{frontmatter}

\blfootnote{MATLAB and Simulink are registered trademarks of The MathWorks,
  Inc. See mathworks.com/trademarks for a list of additional trademarks. Other
  product or brand names may be trademarks or registered trademarks of their
  respective holders.}

\section{Introduction} \label{sec:intro} 

Frequency domain transfer function identification is a widely-used engineering
tool. Commonly this identification task is formulated as a nonlinear least
squares (NLS) problem
(\cite{ljung_system_1999,pintelon_system_2012}). \cite{sanathanan_transfer_1963}
iteration (SK) is a popular approach to solve the NLS problem by fixed-point
iterations, where each iteration is a linear least squares (LLS)
problem. However, traditional implementations of SK iterations with monomial
basis functions suffer from numerical issues, which can be an accuracy
bottleneck.

\cite{gustavsen_rational_1999} introduced the \emph{vector fitting} (VF)
algorithm. VF algorithm is in the SK-iteration framework, but it utilizes
barycentric representation (instead of monomial basis) that is updated at each
iteration step. This approach has improved numerical properties
(\cite{drmac_quadrature-based_2015, drmac_vector_2015}) and many variations
followed. Orthonormal rational basis functions are utilized to further improve
numerics (\cite{deschrijver_orthonormal_2007}). This is called \emph{orthonormal
  vector fitting} (OVF). See \cite{bultheel_orthogonal_2003} for data dependent
orthogonal bases that yield optimal conditioning for SK-iterations.

It is known that SK-iterations may not find a local minimum of the nonlinear
problem when the frequency response measurements contain noise. This issue may
necessitate hard relocation of the identified poles
(\cite{grivet-talocia_improving_2006}), and use of nonlinear optimization
approaches. A recent alternative is the frequency-domain instrumental variable
(IV) iterations (\cite{van_herpen_optimally_2014}). The fixed-points of the
linear IV iterations coincide with the stationary points of the original
nonlinear problem. Hence if convergence is observed, a local minimum is
found. IV iterations have worse condition numbers than SK
iterations. \cite{van_herpen_optimally_2014} introduced bi-orthonormal bases
that yield optimal conditioning for IV iterations to address this issue.

This paper considers continuous-time transfer function estimation from frequency
response measurements using SK and IV iterations with orthonormal rational basis
functions (OVF). These functions lead to zero-pole-gain models, but it is still
possible to perform grey-box transfer function estimation. We also consider
numerical improvements to the OVF algorithm. These improvements are helpful when
data-dependent orthogonal basis functions are not utilized for the sake of
algorithmic simplicity. Our contributions are:
\begin{itemize}
\item An efficient method to enforce constraints on transfer function
  coefficients within the SK and IV iterations with any choice of basis
  functions. This allows incorporating prior knowledge into the estimation
  algorithm such as relative degree, integrators, and affine relationships between
  coefficients.
\item A simple data scaling algorithm, and a domain mapping that maps the
  continuous domain (imaginary axis) to discrete (unit disk). These improve the
  numerics of underlying least squares problems.
\item A minimal state-space realization for the orthonormal rational basis
  functions on the unit disk.
\end{itemize}

The organization of the paper is as follows. Section~\ref{sec:problem} presents
the problem formulation and current approaches. Section~\ref{sec:main} presents
the contributions. Section~\ref{sec:numerical} illustrates the algorithm
performance on two examples. Concluding remarks are in
Section~\ref{sec:conclusion}. The resulting algorithm is implemented in the
\verb"tfest" function (when using frequency-domain data) in System
Identification Toolbox (Release 2016b, for use with MATLAB).

\section{Problem Statement and Current Approaches}
\label{sec:problem} 

Assume that the frequency response measurements of a MIMO system $H(s)$ are
available as $H_1,\ldots,H_l\in\C^{p \times m}$ at frequency points $s_i=j\w_i$
for $i=1,\ldots,l$. The goal is to estimate a MIMO system $H_r(s)$ by
minimizing the cost: \be \label{eq:LSCost} J=\sum_{i=1}^l
\|W(j\w_i)\left(H_r(j\w_i)-H_i\right)\|_F^2 \ee where the degree of $H_r(s)$ is
less than $r\times \min(m,p)$. $W(s)$ is an optional weight. $H_r(s)=N(s)/d(s)$
is a rational function with $d(s):\C^1\to\C^1$ and
$N(s):\C^1\to\C^{p \times m}$.

\cite{sanathanan_transfer_1963} iteration (SK) is a well-known method to
estimate $H_r(s)$ iteratively by utilizing the fact that the problem
\eqref{eq:LSCost} is linear for a fixed denominator. First, rewrite
Equation \eqref{eq:LSCost} exactly: \be J = \sum_{i=1}^l
\frac{|W(j\w_i)|^2}{|d(j\w_i)|^2} \left\| N(j\w_i)-d(j\w_i)H_i \right\|_F^2 \ee
then iterate:
\begingroup\makeatletter\def\f@size{9}\check@mathfonts
\be \label{eq:LSCostIter} J^\kp = \sum_{i=1}^l
\frac{|W(j\w_i)|^2}{|d^\kmp(j\w_i)|^2} \left\| N^\kp(j\w_i)-d^\kp(j\w_i)H_i
\right\|_F^2 \ee
\endgroup
Given $d^\kmp(s)$ from the previous iteration,
computation of the coefficients of $N^\kp$ and $d^\kp$ is a linear least squares
problem. Two common choices for the initial denominator are $d^{(0)}(s)=1$ or
$d^{(0)}(s)$ as a polynomial with lightly damped roots linearly or
logarithmically spaced across $\omega_1$ and $\omega_l$. The problem in
Eq. \eqref{eq:LSCostIter} has a trivial zero solution $d(s)=0$ and $N(s)=0$ that
yields $J^{(k)}=0$. Two approaches to avoid the trivial solution are
constraining the first denominator parameter to 1 or enforcing
$\sum_{i=1}^l Re(d(j\omega_i))=1$ (\cite{gustavsen_improving_2006}).

The key property of the SK fixed-point iterations is as follows: if there is no
noise, and the estimated system order is equal to or greater than the true
order, then the fixed-points of the SK iterations and the stationary points of
the nonlinear problem coincide. Under these assumptions and barring potential
numerical issues, the linear least squares problem in Equation
\eqref{eq:LSCostIter} recovers the true system with $J=0$ in one step,
regardless of the initial choice $d^{(0)}(s)$. Considering these assumptions,
potential issues include:
\begin{enumerate}
\item Numerics: Large condition numbers that may arise in LLS matrices may
  prevent finding accurate solutions. The choice of basis functions for $N(s)$
  and $d(s)$ is critical factor.
\item Fixed-point iterations: When there is measurement noise or data cannot be
  captured by linear models, the fixed-points of SK iterations do not coincide
  with the stationary points of the nonlinear problem. The convergence of SK
  iterations in this case is an open problem. Even if the iterations converge,
  the result is not necessarily a local optima. IV iterations is an alternative
  approach whose fixed-points coincide with the stationary points of the
  nonlinear problem.
\item Incorrect model order: This leads to the same issues in (2), but has the
  partial remedy of being a tunable parameter at estimation time. Model order
  estimation in SK and IV iteration framework is an open problem, but some
  promising results do exist (\cite{grivet-talocia_improving_2006,
    drmac_vector_2015}).
\end{enumerate}

The solution found for the NLS problem in Equation \eqref{eq:LSCost} depends on
all the above factors. Any of these factors can constitute an accuracy
bottleneck depending on the specific data. Sections \ref{sec:numericsAndVF} and
\ref{sec:IVIterations} provide details of potential issues with numerics as well
as SK and IV fixed-point iterations, respectively.

\subsection{Basis Functions  and Vector Fitting}
\label{sec:numericsAndVF}

The condition numbers of the least squares matrices that arise in SK iterations
depend on the basis function choice and the weight
$\frac{W(s)}{d^{(k-1)}(s)}$. The basis function choice involves algorithmic
complexity versus accuracy trade-off. Some common basis functions and a brief
discussion of this trade-off are presented here.

A generic representation for the rational function $H_r(s)$ is:
\be \label{eq:baryOVF} H_r^\kp=\frac{N^\kp(s)}{d^\kp(s)}=\frac{\sum_{\nu=0}^r
  N_\nu^\kp B_\nu^\kp(s)}{\sum_{\nu=0}^r d_\nu^\kp B_\nu^\kp(s)} \ee where
$B_\nu^{(k)}(s):\C^1\to\C^1$ is the $\nu^{th}$ basis function
utilized in iteration k. A simple implementation with monomials $B_\nu(s)=s^\nu$
leads to \be
H_r^\kp=\frac{N^\kp(s)}{d^\kp(s)}=\frac{N_0^\kp+\sum_{\nu=1}^{r}N_\nu^{(k)}\,s^\nu}{d_0^\kp+\sum_{\nu=1}^{r}d_\nu^{(k)}\,s^\nu}\ee
Monomial basis leads to large condition numbers. The vector fitting approach
(\cite{gustavsen_rational_1999}) improves the numerics on two fronts. First, the
$N^{(k)}(s)$ and $d^{(k)}(s)$ are expressed in barycentric form with
$B^{(k)}_\nu(s)=\frac{1}{s-\lambda^{(k)}_\nu}$ where $\lambda^{(k)}_\nu\in\C^1$
is the $\nu^{th}$ distinct interpolation point: \be \label{eq:baryVF}
H_r^\kp=\frac{N^\kp(s)}{d^\kp(s)}=\frac{N_0^\kp+\sum_{\nu=1}^{r}
  \frac{N_\nu^\kp}{s-\lambda_\nu^\kp}}{d_0^{(k)}+\sum_{\nu=1}^{r}
  \frac{d_\nu^\kp}{s-\lambda_\nu^\kp}} \ee This yields improved condition
numbers over monomials. Second, $\lambda^{(k)}_\nu$ are chosen as the zeros of
the identified $d^{(k-1)}(s)$ in the previous iteration step. This elegant
choice eliminates the need for the weight $\frac{1}{d^{(k-1)}(s)}$, which
typically worsens the condition numbers iterations evolve. These two ideas in
the VF algorithm lead to better numerical stability (\cite{drmac_vector_2015}).

The VF algorithm is not much more complicated than using monomial basis
functions, yet it is significantly better regarding numerics. This makes it a
good candidate for many applications. However, problems do exist where this
method is not sufficient. One example is systems with closely packed or repeated
poles. To see this, note that the VF algorithm uses the identified poles as the
interpolation points $\lambda^{(k)}$ for barycentric representation. Repeated
$\lambda^{(k)}$ values yield linearly dependent basis functions.

Two further basis function candidates for handling such cases are highlighted
here. One is orthonormal rational polynomials obtained by analytical
Gram-Schmidt orthogonalization of barycentric basis functions
(\cite{akcay_orthonormal_1999,ninness_basisfcns_1997}). These typically yield
better condition numbers, and do not suffer from numerical issues with close or
repeated poles (except when there are multiple integrators). The VF algorithm
with such bases is called OVF. Orthonormal rational polynomials on the unit disk
are utilized in \verb"tfest", and explained further in Section
\ref{sec:ovfSSRealization}. Second option is data dependent orthogonal
bases. See \cite{bultheel_orthogonal_2003}) for matrices observed in SK
algorithm, and \cite{van_herpen_optimally_2014} for matrices in the IV
algorithm. These data dependent orthogonal bases ensure that the least squares
matrix have a condition number 1. The drawback is the higher computational
complexity of the construction of these bases compared to the VF and OVF
approaches. This is a future improvement direction for the \verb"tfest"
implementation.

\subsection{Instrumental Variable Iterations}
\label{sec:IVIterations}

In the presence of noise, nonlinearities or wrong model order choice, the
fixed-points of SK iterations do not coincide with the stationary points of
the nonlinear estimation problem. \cite{whitfield_asymptotic_1987} shows that
even when the SK iterations converge to a solution, the solution is not
necessarily a local optima.

An alternative is the so-called instrumental-variable (IV) iterations. IV
iterations have the property that its fixed-points coincide with the stationary
points of the nonlinear least squares problem, even in presence of the
aforementioned conditions (\cite{van_herpen_optimally_2014}). There is no
convergence guarantee, but if convergence is observed, the result is a local
optima. IV iterations aim to find the stationary points (i.e. where derivative
with respect to parameters ($N_\nu^\kp$, $d_\nu^\kp$) are zero) of the nonlinear
problem directly instead of directly trying to minimize the cost function in
Equation \eqref{eq:LSCost}. The new cost function minimized in IV iterations is:
\begingroup\makeatletter\def\f@size{8}\check@mathfonts
\begin{multline*} 
J_{iv} = \sum_{i=1}^l |W(j\w_{i})|^2\ \sum_{\nu=0}^r
\left(\left\|\frac{\partial \tilde{J}_i}{\partial N_\nu}\right\|_F^2+
\left\|\frac{\partial \tilde{J}_i}{\partial d_\nu}\right\|_F^2\right)
\end{multline*}
\endgroup
 where $\tilde{J}_i(j\w)=(H_r^\kp(j\w)-H_i)(H_r^\kp(j\w)-H_i)^T$
\begingroup\makeatletter\def\f@size{8}\check@mathfonts
\begin{equation*}
 \begin{aligned}
 \frac{\partial \tilde{J}_i}{\partial N_\nu}&= -2 \left(H_r^\kp(j\w_i)-H_i\right)\left(\frac{\partial H_r^\kp(j\w_)}{\partial N_\nu}\right)^T \\
 \frac{\partial \tilde{J}_i}{\partial d_\nu}&= -2
 \left(H_r^\kp(j\w_i)-H_i\right)\left(\frac{\partial H_r^\kp(j\w_i)}{\partial
 d_\nu}\right)^T
\end{aligned}
\end{equation*}
\endgroup
This is a NLS problem due to nonlinear dependence of $H_r^\kp$ and its partial
derivatives on denominator coefficients. The terms that appear in the
denominator of the nonlinear cost function is replaced by the estimates from the
previous iteration, similar to SK iterations. This again yields linear least
squares problems for each iteration step.

The IV iterations can be utilized standalone for solving the nonlinear
optimization problem in Eq. \eqref{eq:LSCost}, as in
\cite{van_herpen_optimally_2014}. The implementation in \verb"tfest" uses IV
iterations in succession to the SK iterations. This is to find a solution near a
reasonable local optima first with the SK iterations, which is then utilized for
initializing the IV iterations. This is a conservative approach given that IV
iterations are frequently successful at good local optima points. This
conservative choice was done to increase the chance of finding good solutions
given that \verb"tfest" has a wide user base and it is used for a large variety
of datasets.

\section{Contributions} \label{sec:main}

\subsection{Domain Mapping}
\label{sec:domainMapping}

Numerical issues with continuous-time transfer function estimation are most
prevalent when the measurement points $s_i=j\omega_i$ span a wide frequency
range. The large magnitude variations of $s_i$ are observed across the
rows of the least squared matrices, and lead to numerical issues.


It is possible to perform the estimation on a different domain, and transform
the final result back. This idea, for instance in the form of scaling the
$s$-domain (\cite{pintelon_scaling_2005}), is found in the early continuous-time
transfer function estimation literature where monomial bases are commonly
used. More recent literature (VF, OVF) typically do not perform such domain
mapping, and instead just scale the columns of the least squares matrices at
each iteration step. This is because column scaling, combined with the use of
barycentric or orthonormal rational polynomials, yields sufficient condition
numbers on a wider range of estimation problems.

Any invertible mapping can be utilized, and the bilinear mapping
$q(s)=\frac{\alpha+s}{\alpha-s}$ is proposed here. This idea is from
\cite{balas_2002}. Bilinear mappings provide more flexibility over simple
scaling, while maintaining a linear relationship between the transfer function
coefficients in the original and transformed domains. The latter property is
utilized for efficient handling of parameter constraints discussed in Section
\ref{sec:ParameterConstraints}. The mapping $q(s)$ maps the halfplane
$s=j\omega$ for $\omega\in[0,\infty)$ to the upper half of the unit circle. The
scalar $\alpha\in(0,\infty)$ is a design parameter. The mapped points in $q$
domain have unit magnitude. This eliminates the magnitude variations across the
basis function rows observed in $s$ domain. An idea for $\alpha$ is to maximize
the distance between the endpoints in $s$-domain through the transformation
$q(s)$. Precisely, $\argmax_\alpha |q(j\omega_1)-q(j\omega_l)|$. The unique
stationary point of this problem is $\alpha=\sqrt{\omega_1\omega_l}$.


Effect of the bilinear domain mapping on the matrix condition numbers is tested
with barycentric basis functions (VF) for a single SK iteration step in Equation
\eqref{eq:LSCostIter}. The first denominator parameter is fixed to 1. The data
$H_i(j\omega_i)$ is from the $20^{th}$ order model in Section
\ref{sec:experimentsUnconstrained}. The weight $W(j\omega_i)$ and
$d^{(k-1)}(j\omega_i)$ is set to 1. The interpolation points $\lambda_i$ are set
as the true poles of the model. For domain mapping, the interpolation points are
also mapped as $\bar{\lambda}_i=q(\lambda_i)$. These interpolation points
represent a good candidate for the final SK-iteration step, hence the observed
condition numbers have a direct impact on the final accuracy.  The barycentric
bases $1/(s-\lambda_i)$ yield the condition number
$4.9{\cdot}10^{8}$. Barycentric bases in transformed domain
$1/(q-\bar{\lambda}_i)$ yield $9.1{\cdot}10^{5}$.

VF literature typically suggests scaling the columns of the least squares
matrices by Euclidean norm. There are also suggestions against this scaling
(\cite{drmac_vector_2015}) because it may amplify the impact of
noise. \verb"tfest" does not use column scaling, but the impact of the column
scaling is also tested. This scaling reduces the condition number with bases in
original domain $1/(s-\lambda_i)$ to $1.2{\cdot}10^{3}$, and in the
transformed domain $1/(q-\bar{\lambda}_i)$ to $8.1{\cdot}10^{1}$. These show
that the basis functions in the transformed domain $q$ are better conditioned
with or without column scaling.


\subsection{Grey-Box Estimation: Parameter Constraints}
\label{sec:ParameterConstraints}

Many grey-box transfer function estimation scenarios involve fixed or bounded
numerator or denominator coefficients, for instance known integrators, relative
degree or bounds on a mass-spring-damper system parameters.

It is straightforward to handle affine constraints within the SK iteration
framework when the numerators and the denominator are expressed in terms of
monomials. Then the estimated parameters are the transfer function coefficients
themselves. This corresponds to solving a constrained least-squares problem in
each iteration step, instead of an unconstrained one.

When barycentric representation (VF) or orthonormal rational basis functions
(OVF) are used, the estimated parameters are no longer the transfer function
coefficients. The key point is that there is a linear relationship between the
estimated parameters and the transfer function coefficients. Here a numerical
method is presented to calculate this relationship for barycentric or
orthonormal rational basis functions. Domain transformation (Section
\ref{sec:domainMapping}) is also accounted for in this treatment.

Make the following three assumptions without loss of generality to ease the
notational burden: the transfer function is SISO, there are constraints only on
the denominator coefficients, and barycentric representation (VF) is
utilized. Let $d(s)=\sum_{\nu=0}^rd_{\nu}s^\nu$ be the denominator of the sought
after transfer function. Let $d(q)=d_{q0}+\sum_{\nu=1}^{r}d_{q\nu}B_\nu(q)$ be the
estimated denominator, in domain $q=\frac{\alpha+s}{\alpha-s}$, with basis
functions $B_\nu(q)=\frac{1}{q-\lambda_\nu}$. The aim is to find the linear
relationship between transfer function coefficients $d_{\nu}$ and estimated
coefficients $d_{q\nu}$ for $\nu=0,1,...,r$.
\begingroup\makeatletter\def\f@size{8}\check@mathfonts
\begin{align}
d(q) & = d_{q0}+\sum_{\nu=1}^r d_{q\nu}\, B_{\nu}(q) \label{eq:paramConstQ} \\ \nonumber
     & = \frac{\sum_{\nu=0}^r \bar{d}_{q\nu} \, q^\nu}{\prod_{\nu=1}^r
       (q-\lambda_\nu)} \\ \nonumber
     & = \frac{\sum_{\nu=0}^r \bar{d}_{q\nu} \, (\alpha+s)^\nu \, (\alpha-s)^{r-\nu}}
     {(\alpha-s)^r \, \prod_{\nu=1}^r  (q-\lambda_\nu)} \\ 
     & = \frac{\sum_{\nu=0}^r d_{\nu} \, s^\nu}{(\alpha-s)^r \, \prod_{\nu=1}^r (q-\lambda_\nu)} \label{eq:paramConstZ}
\end{align}
\endgroup
Here $\bar{d}_{q\nu}$ are some intermediate variables. The right-hand side of
Equation \eqref{eq:paramConstQ} contains the estimated parameters $d_{q\nu}$,
and the numerator of the right-hand side of Equation \eqref{eq:paramConstZ}
contains the transfer function coefficients of interest $d_\nu$. Both quantities
appear linearly in the respective equations. This linear mapping can be
extracted by evaluating these equations at $r+1$ points:
\begingroup\makeatletter\def\f@size{8}\check@mathfonts
\begin{align}
\nonumber
\begin{bmatrix}
 \vdots & \vdots & \vdots & \vdots  \\
 1      & B_1(q) & \cdots & B_r(q) \\  
 \vdots & \vdots & \vdots & \vdots 
\end{bmatrix}
\begin{bmatrix}
 d_{q0} \\ 
 \vdots \\
 d_{qr}
\end{bmatrix} = \\
\begin{bmatrix}
 \ddots &   0    & 0  \\
 0      & \frac{1}{(\alpha-s)^r \, \prod_{\nu=1}^r (q-\lambda_\nu)} & 0  \\  
 0      & 0 & \ddots
\end{bmatrix}
\begin{bmatrix}
 \vdots & \vdots & \vdots & \vdots  \\
 1      & s      & \cdots & s^r      \\  
 \vdots & \vdots & \vdots & \vdots 
\end{bmatrix}
\begin{bmatrix}
 d_{0} \\ 
 \vdots \\
 d_{r}
\end{bmatrix} \label{eq:paramConstMatS}
\end{align}
\endgroup
The aim is to solve this equation for $[d_0 \dots d_r]^T$. The first matrix on
the right hand side of \eqref{eq:paramConstMatS} is diagonal, hence easy to
invert.  $[1\; s \cdots s^r]$ is a Vandermonde matrix. Choose the evaluation
points $s$ uniformly spaced on the unit disk to make the Vandermonde matrix
unitary (after scaling with $1/\sqrt{r+1}$), which is then also easy to invert. 

The constructed linear constraints can be utilized in SK iterations without
modifying the cost function in Equation \eqref{eq:LSCostIter}. Each iteration
step is a constrained LLS problem in this case. For IV iterations, the cost
function also needs to be modified because stationary points of the NLS problem
can be at points where derivative is not zero. The method of Lagrange
multipliers can be used to extend the IV iteration cost function to find the
stationary points of the constrained nonlinear problem
(\cite{bertsekas_nonlinear_1999}).

\subsection{Measurement Scaling}
\label{sec:DataScaling}

The measured frequency response $H_1,\dotsc,H_l\in\C^{p \times m}$ enters the SK
and IV iteration linear least-squares matrices. Therefore, the choice of
measurement units has an impact on the condition number of the least-squares
matrices, and in turn the final fit quality. This is important when using
monomial, barycentric (VF) or orthonormal rational polynomial (OVF) basis
functions.


Let $B^{(k)}=[B_1^{(k)} \dots B_r^{(k)}]$ be the basis function matrix at
iteration $k$. For a given input-output channel $(i,j)$, the corresponding rows
of the least squares matrices contain $B^{(k)}$ once unscaled, and once
row-scaled by measurements $H(i,j)$. A heuristic is to balance the row-scaling
induced by $H(i,j)$ around 1 by scaling
$c(i,j)=\sqrt{\max_t|H_t(i,j)| \min_t|H_t(i,j)|}$, which yields scaled
measurements $\bar{H}(i,j)=c(i,j)H(i,j)$.

This magnitude scaling is tested on the example in Section
\ref{sec:experimentsUnconstrained}. This scaling reduces the worst condition
number observed during the SK and IV iterations from $4.2{\cdot}10^{14}$ to
$2.6{\cdot}10^{12}$. This is a modest improvement due to good scaling of the
original data. Nevertheless, measurement scaling is straightforward,
computationally cheap and useful.

\subsection{State-Space Realization for Orthonormal Rational Basis Functions}
\label{sec:ovfSSRealization}

The orthonormal rational basis functions on the unit disk are due to
\cite{ninness_basisfcns_1997}. Let $\xi_\nu\in\C^1$ be the $\nu^{th}$ interpolation
point where $\nu=1,2,\ldots,r$. These basis functions have the form in Eq.
\eqref{eq:orthBasisReal} if $\xi_\nu$ is real, and the form in Eq.
\eqref{eq:orthBasisCplx} if $(\xi_\nu,\xi_{\nu+1})$ are complex-conjugate pairs.
\begin{equation}
\label{eq:orthBasisReal}
B_\nu(q) = \frac{\sqrt{1-|\xi_\nu|^2}}{q-\xi_\nu} \prod_{\nu=1}^{n-1} \frac{1-\xi_\nu^*q}{q-\xi_\nu}
\end{equation}
\begin{equation}
\begin{aligned}
\label{eq:orthBasisCplx}
B_\nu(q) = \frac{\sqrt{1-|\xi_\nu|^2}(\beta_1q+\mu_1)}{q^2-2Re(\xi_\nu)+|\xi_\nu|^2} \prod_{\nu=1}^{n-2}
\frac{1-\xi_\nu^*q}{q-\xi_\nu}\\
B_{\nu+1}(q) = \frac{\sqrt{1-|\xi_\nu|^2}(\beta_2q+\mu_2)}{q^2-2Re(\xi_\nu)+|\xi_\nu|^2} \prod_{\nu=1}^{n-2}
\frac{1-\xi_\nu^*q}{q-\xi_\nu}
\end{aligned}
\end{equation}
Here $B_n:\C^1\to\C^1$ is the $n^{th}$ basis function,
$Re(\cdot)$ is the real part of a complex number. There are infinite number of
choices for $\beta_1$, $\beta_2$, $\mu_1$, $\mu_2$
(\cite{ninness_basisfcns_1997}). Each SK and IV iteration step estimate a
denominator polynomial $d^{(k)}(q)=d_0+\sum_{\nu=1}^rd_\nu B_\nu(q)$, then
extract its zeros to find the interpolation points $\xi_\nu$ for the next
iteration. A minimal state-space realization of $d^{(k)}(q)$ is needed for this
operation. A state-space realization for orthonormal rational polynomials on the
imaginary axis is in \cite{deschrijver_orthonormal_2007}. Here a similar
construction is presented for orthonormal rational polynomials defined on the
unit disk.
\begin{figure}[h!]
\setlength{\unitlength}{1pt}
\centering
\begin{picture}(210, 31)(0,0)
\multiput(0,20)(37,0){6}{\vector(1,0){7}}
\put(7,10){\framebox(30,21){$ \frac{1-\xi_1^*q}{q-\xi_1}$}}
\put(12,0){$\scriptstyle G_1(q)$}
\put(44,10){\framebox(30,21){$ \frac{1-\xi_2^*q}{q-\xi_2}$}}
\put(49,0){$\scriptstyle G_2(q)$}
\put(91,17){$\cdots$}
\put(118,10){\framebox(30,21){$ \frac{1-\xi_\nu^*q}{q-\xi_\nu}$}} 
\put(123,0){$\scriptstyle G_\nu(q)$}
\put(165,17){$\cdots$}
\put(192,10){\framebox(30,21){$ \frac{1-\xi_r^*q}{q-\xi_r}$}}
\put(198,0){$\scriptstyle G_r(q)$}
\end{picture}
\caption{Components of the basis functions $B_n(q)$}
\label{fig:cascadeIC}
\end{figure}
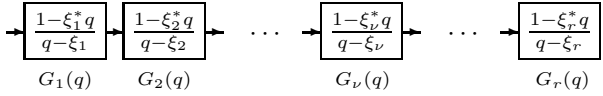

\vspace{-.1in}
Denote the state-space model matrices to be constructed as $(A,B,C,D)$. The idea
is to choose $(A,B)$ such that the states $(qI-A)^{-1}B$ correspond to the basis
functions $[B_1(q) B_2(q) \dots B_r(q)]^T$, except the $\sqrt{1-|\xi_\nu|^2}$
scalings. Consider the cascade connection in Fig. \ref{fig:cascadeIC} for this
purpose. Realize the $n^{th}$ component is with $(a_n,b_n,c_n,d_n)$ such that
its output is $\frac{1-\xi_\nu^*q}{q-\xi_\nu}$, and its state is $\frac{1}{{q-\xi_\nu}}$.
Then the series connection of all components per Equation \eqref{eq:ovfSSAB} has
its $\nu^{th}$ state as $B_\nu(q)$ (except the scaling $\sqrt{1-|\xi_\nu|^2}$).

If $(\xi_\nu,\xi_{n+1})$ are complex-conjugate pairs, $G_n(q)G_{n+1}(q)$ must be
realized together in order to have a realization with real coefficients. In this
case the output is
$\frac{|\xi_\nu|^2q^2-2Re(\xi_\nu)+1}{q^2-2Re(\xi_\nu)+|\xi_\nu|^2}$, and the
states are
$\frac{1}{{q^2-2Re(\xi_\nu)+|\xi_\nu|^2}}[\beta_1q+\mu_1 \;\;
\beta_2q+\mu_2]^T$. A realization that fits these requirements is in Table
\ref{table:ssRealization}. This specific realization is obtained in three
steps. First, fix $b_v=[\beta_1\;\beta_2]^T$ and $d_v=|\xi_\nu|^2$ to avoid
over-parametrization. Second, solve
$(qI-a_v)^{-1}b_v=\frac{1}{{q^2-2Re(\xi_\nu)+|\xi_\nu|^2}}[\beta_1q+\mu_1 \;
\beta_2q+\mu_2]$ for the elements of $a_v$. Finally, solve
$c_v(qI-a_v)^{-1}b_v+d_v=\frac{|\xi_\nu|^2q^2-2Re(\xi_\nu)+1}{q^2-2Re(\xi_\nu)+|\xi_\nu|^2}$
for the elements of $c_v$.

The matrices $(C,D)$ capture the scalings $\sqrt{1-|\xi_\nu|^2}$ and the estimated
parameters $d_0,\dots,d_r$. Specifically, $D=d_0$ and $\nu^{th}$ element of
$C\in\R^{1 \times r}$ is $d_\nu\sqrt{1-|\xi_\nu|^2}$.

\begin{table}[ht]
\centering
\begin{tabular}{ | l | l | }
    \hline
    $\xi_\nu$ real & If $\xi_\nu\neq0$, $(a_v,b_v,c_v,d_v)=(\xi_\nu,1,1-\xi_\nu^2,-\xi_\nu)$  \\ \cline{2-2}
                 & If $\xi_\nu=0$, $(a_v,b_v,c_v,d_v)=(0,1,1,0)$    \\    \hline
    $(\xi_\nu, \xi_{v+1})$ &  $a_v\in\R^{2\times2},b_v=[\beta_1\; \beta_2]^T,c_v\in\R^{1\times2},d_v=|\xi_\nu|^2$  \\ 
     complex pair &
                    $a_v(1,1)=\frac{|\xi|^2+(\mu_1\mu_2)/(\beta1\beta2)+(2Re(\xi_\nu)\mu_2)/\beta_2}{\mu_2/\beta_2-\mu_1/\beta_1}$ \\
                  & $a_v(1,2)=\frac{\mu_1+2\beta_1Re(\xi_\nu)-\beta_1a_v(1,1)}{\beta_2}$ \\
                  & $a_v(2,1)=\frac{\mu_2+\beta_2a_v(1,1)}{\beta_1}$ \\ 
                  & $a_v(2,2)=2Re(\xi_\nu)-a_v(1,1)$ \\

                  & $c_v(1,1)=\frac{2\mu_2Re(\xi_\nu)(|\xi|^2-1)+(|\xi_\nu|^4-1)\beta_2}{\beta_1\mu_2-\mu_1\beta_2}$ \\
                  & $c_v(1,2)=\frac{2Re(\xi_\nu)(|\xi_\nu|^2-1)-c_v(1,1)\beta_1}{\beta_2}$\\
    \hline
    \end{tabular}
\caption{Realizations of components of interconnection in Fig. \ref{fig:cascadeIC}}
\label{table:ssRealization}    
\end{table}
\begin{equation}
\begin{aligned}
\label{eq:ovfSSAB}
A&=\begin{bmatrix}
  a_1                   & 0         & \cdots & 0 \\
  b_2c_1                & a_2       & \cdots & 0 \\
  b_3d_2c_1             & b_3c_2     & \cdots  & 0 \\
  b_4d_3d_2c_1          & b_4d_3c_2  & \cdots   & \vdots \\
  \cdots                & \cdots   &  \cdots  & 0 \\
  b_kd_{k-1} \cdots d_2c_1 & b_kd_{k-1} \cdots d_3c_2  & \cdots   & a_k \\
\end{bmatrix} \\
B&=[b_1 \;\;\; b_2d_1 \;\;\; b_3d_2d_1 \cdots \;\;\; b_kd_{k-1}\dots d_1]^T
\end{aligned}
\end{equation}

\subsection{Algorithm Summary}
The algorithm in \verb"tfest" can be summarized as:
\begin{enumerate}
\item Map $s$ domain to $q$ via $q(s)=\frac{\alpha+s}{\alpha-s}$
\item Scale measurements $H_i$ (Section \ref{sec:DataScaling}).
\item Initial fit: Use monomial basis with $d^{(0)}(q)=1$. 
\item SK iterations: Use orthonormal rational polynomial basis functions on unit
  disk. Iterate until maximum number of iterations, or convergence. Update basis
  functions at each step.
\item IV iterations: Use the final set basis functions used  in SK
  iterations. Iterate until maximum number of iterations, or convergence.
\item Use the best solution found for the NLS problem throughout all steps
  (initial fit, SK and IV iterations). Calculate the corresponding zero-pole-gain
  model.
\item Revert $s$ to $q$ domain mapping via $s=\frac{\alpha(q-1)}{q+1}$.
\item Revert measurement scaling.
\item Convert zero-pole-gain to transfer function model.
\end{enumerate}

\section{Numerical Experiments} \label{sec:numerical} 

The \verb"tfest" command in System Identification Toolbox (Release 2016b, for
use with MATLAB) is used for the experiments. The results compare very favorably
with previous releases and other existing algorithms for many frequency domain
datasets. Two experiments are presented here due to space constraints.

\subsection{Experiment Without Parameter Constraints}
\label{sec:experimentsUnconstrained}

A randomly generated model was used for the experiment:
\begin{equation}
G(s) = \sum_{k=1}^{10} \frac{r_k\;\omega_k^2}{s^2 + 2\zeta_k\omega_ks + \omega_k^2}
\end{equation}
where the parameters $r_k$, $\omega_k$, $\zeta_k$ are randomly generated from a
predetermined range. This is a dynamically rich model: it contains ten lightly
damped modes that are spread over seven decades. The frequency response of
$G(s)$ was extracted at 700 logarithmically spaced points in $[10^{-1},10^{6}]$
$rad/s$. Multiplicative noise was added as $G(j\omega_i)(1+\epsilon(j\omega_i))$
where $\epsilon\in\C^1$ is $\epsilon(j\omega_i)=n_1e^{jn_2}$. $n_1$ and $n_2$
are zero mean Gaussian random numbers with variance 0.01, so the signal-to-noise
ratio is approximately $20\,dB$.

\begin{figure}[ht]
\centering
 \includegraphics[width=0.46\textwidth]{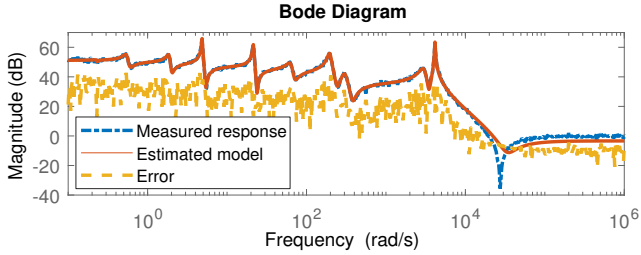}
 \caption{$20^{th}$ order black-box model estimate}
\label{fig:blackBoxExample}
\end{figure}

Figure \ref{fig:blackBoxExample} presents the measurements and the
$20^{th}$ order model estimate. The model captures the dynamics of interest, despite
the high model order for transfer functions. Only the valley near
$2.74\cdot10^4\;rad/s$ was missed. The main reason is the limited contribution
of the small magnitude data at and around the valley to the cost function. 
Using a frequency-based weight $W(s)$ is a straightforward remedy to this
problem, which is available through the WeightingFilter option of the
\verb"tfestOptions" command.

\subsection{Experiment with Parameter Constraints}
\label{sec:experimentsConstrained}

The model $G(s)$ in Equation \eqref{eq:constrainedModel} is used for the
experiment. $G(s)$ has an integrator and a relative degree 3. Three leading
numerator coefficients and the last denominator coefficient are fixed to 0
during estimation. The frequency response is extracted at 300 logarithmically
spaced points in $[1,10^4]$ $rad/s$. Multiplicative noise is added as
explained in Section \ref{sec:experimentsConstrained}, to have a $20dB$
signal-to-noise ratio.
\begin{equation}
\label{eq:constrainedModel}
\small
G(s) = \frac{1.2{\cdot}10^8\,s^2 + 4.8{\cdot}10^7\,s + 5{\cdot}10^{10}}{s^5 + 200\,s^4 +
  4{\cdot}10^6\,s^3 + 1.7{\cdot}10^6\,s^2 + 1.6{\cdot}10^9\,s}
\end{equation}

\begin{figure}[ht]
 \centering
 \includegraphics[width=0.46\textwidth]{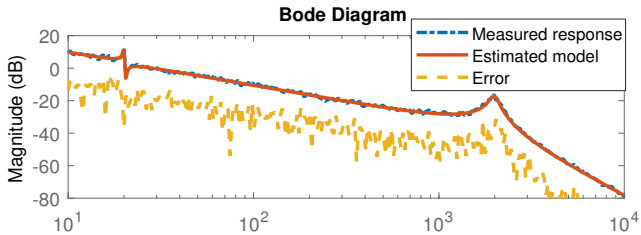}
 \caption{Measured response and grey-box model estimate}
\label{fig:constrainedFitBode}
\end{figure}

Figure \ref{fig:constrainedFitBode} shows the Bode magnitude plot for the noisy
measurements and the fitted model. The estimated model in Equation
\eqref{eq:constrainedModelEst} captures system dynamics well. The approach in 
Section \ref{sec:ParameterConstraints}
successfully enforces the poles and zeros in transformed domain $q$ to honor the
transfer function coefficient constraints in $s$ domain.
\begin{equation}
\label{eq:constrainedModelEst}
\small
G_{est}(s) = \frac{1.17{\cdot}10^8\,s^2 + 3.32{\cdot}10^7\,s + 4.91{\cdot}10^{10}}{s^5 + 211\,s^4 +
  4.0{\cdot}10^6\,s^3 + 1.4{\cdot}10^6\,s^2 + 1.6{\cdot}10^9\,s}
\end{equation}

\section{Concluding Remarks} \label{sec:conclusion}

The SK and IV iterations are commonly implemented with monomial, barycentric or
orthonormal rational polynomial basis functions on the imaginary
axis. Transforming the problem domain onto the unit disk frequently leads to
important numerical improvements. A simple scaling of the measurement data is
also helpful.

Even though use of barycentric or orthonormal rational polynomials lead to
zero-pole-gain models, the numeric benefits of these bases can still be utilized
for grey-box transfer function estimation. The linear relationship between these
basis functions and transfer function coefficients can be efficiently calculated
by utilizing the properties of the Vandermonde matrices.

\begin{ack}
  We would like to acknowledge and thank Professor Peter J. Seiler for sharing his
  expertise with us, which was instrumental for this work.

  We would also like to acknowledge and thank Professor Lennart Ljung for his
  continuous support, feedback and encouraging comments which motivated us and
  improved the quality of this work.
\end{ack}

\bibliography{lsfitfrd}
\end{document}